\pdfoutput=1
\documentclass[aps,pra,reprint,superscriptaddress,longbibliography,floatfix]{revtex4-2}

\usepackage[T1]{fontenc}
\usepackage[utf8]{inputenc}
\usepackage{amsmath,amssymb,amsfonts,bm,mathtools}
\usepackage{amsthm}
\usepackage{graphicx}
\usepackage{booktabs}
\usepackage{multirow}
\usepackage{hyperref}
\hypersetup{hidelinks}
\usepackage{array}

\newtheorem{definition}{Definition}
\newtheorem{remark}{Remark}

\begin{document}

\title{Free-Space CV-QKD with Single-Mode Fiber Reception:\\ Effective Coupling Statistics and Protocol-Dependent Reference Noise}

\author{Hesham S. Ibrahim}
\author{Arnaud Coatanhay}
\affiliation{Lab-STICC, UMR CNRS 6285, ENSTA, Institut Polytechnique de Paris,\\
2 rue Fran\c{c}ois Verny, 29806 Brest Cedex 9, France}

\begin{abstract}
We study free-space continuous-variable quantum key distribution (CV-QKD) with single-mode fiber (SMF) reception under atmospheric turbulence. The optical channel is modeled by split-step propagation through random phase screens, followed by finite-aperture collection and projection onto the guided receiving mode. We first examine the standard GG02 setting and ask which receiver-side observable is sufficient for effective key-rate prediction. We show that a mean-loss description is generally too optimistic, whereas a scalar effective law for the SMF coupling efficiency provides an accurate downstream Gaussian-channel description within the effective model considered here. We then extend the optical model to a pilot-assisted architecture in which the signal and pilot propagate through correlated but non-identical turbulent realizations generated by a frozen-flow construction. In this case, the signal coupling law alone is no longer sufficient: signal--pilot phase mismatch and loss of post-coupling coherence produce an additional protocol-dependent reference-noise penalty. The results distinguish two regimes: a scalar coupling description is largely adequate for GG02, while transmitted-reference architectures require an additional differential reference observable beyond the signal coupling statistics.
\end{abstract}
\maketitle

\section{Introduction}
\label{sec:introduction}

Continuous-variable quantum key distribution (CV-QKD) based on coherent states and coherent detection is a major platform for quantum-secure communications because it is compatible with standard telecom technology and admits a well-developed Gaussian-state security framework \cite{Grosshans2002,Grosshans2003,GarciaPatron2006,Weedbrook2012,Diamanti2015,Leverrier2013,Leverrier2015,Leverrier2017,Laudenbach2018,Pirandola2020}. Integrated photonic implementations further support the technological maturity of the CV-QKD platform \cite{Zhang2019}.
In this context, the Gaussian-modulated coherent-state GG02 protocol remains the natural reference model for effective channel descriptions and performance studies \cite{Grosshans2002,Grosshans2003,GarciaPatron2006,Leverrier2013,Leverrier2015,Laudenbach2018}.

Beyond fiber links, free-space CV-QKD is relevant for urban line-of-sight channels, mobile optical links, and satellite scenarios in which guided transmission is unavailable or impractical \cite{Usenko2012,Heim2014,Wang2018,Shen2019,Dequal2021,Hosseinidehaj2019,Hosseinidehaj2021,Pirandola2021,Derkach2024}. In such systems, atmospheric turbulence acts on the optical field before coherent detection. It affects not only the average received power, but also the spatial structure of the field through beam wander, scintillation, wavefront distortion, and mode degradation \cite{Andrews2005,Schmidt2010,Vasylyev2012,Vasylyev2016,Vasylyev2017}. Recent work has also examined adaptive-optics mitigation in free-space CV-QKD \cite{Sayat2026}. This raises a basic modeling question: once turbulent propagation and coherent reception are described explicitly, what is the physically relevant effective channel variable for CV-QKD?

This question becomes particularly sharp when the receiver includes single-mode fiber (SMF) coupling. In that case, the detected signal is not determined solely by the power intercepted by the receiver aperture. It also depends on the overlap between the distorted received field and the guided receiving mode \cite{Dikmelik2005}. The receiver therefore performs a modal filtering operation, and the effective transmission seen by the QKD protocol is more naturally described by a coupling observable than by aperture-collected power alone \cite{Dikmelik2005,Vasylyev2016,Vasylyev2018}.

A number of free-space CV-QKD studies describe atmospheric propagation through fluctuating transmittance models, probability distributions of transmittance, or related fading descriptions \cite{Vasylyev2012,Usenko2012,Heim2014,Vasylyev2016,Vasylyev2018,Wang2018,Chai2019,Ruppert2019,Hosseinidehaj2021}. Such models are often effective for performance analysis, but they do not by themselves identify which receiver-side observable should be retained once the full optical propagation and the SMF projection are treated explicitly. This is the first question addressed in the present work.

Our first objective is therefore to determine the appropriate effective description of a turbulent free-space link with SMF reception in the GG02 setting. Starting from a wave-optics propagation model followed by finite-aperture collection and projection onto the guided mode, we ask whether the full propagation--reception chain can be reduced to a scalar effective law for the SMF coupling efficiency, or whether a more structured description is required. To answer this, we compare a mean-loss approximation, moment-based effective Gaussian closures, and scalar surrogate laws inferred from the full optical simulations.

Our second objective is to examine when the protocol itself becomes physically relevant at the channel-model level. If the effective channel is fully described by the scalar statistics of the signal coupling, then different protocols mainly process the same scalar channel in different ways. By contrast, a transmitted-reference or pilot-assisted architecture may depend on additional observables beyond the signal coupling alone. In such a setting, the atmosphere may introduce not only amplitude fading, but also a differential signal--reference mismatch, which then appears as an additional effective noise contribution at the coherent receiver \cite{Qi2015,Soh2015,Marie2017,Laudenbach2019}.

To probe this regime, we extend the optical model to a pilot-assisted configuration in which the signal and pilot propagate through correlated but non-identical turbulent realizations generated by a frozen-flow construction \cite{Taylor1938,Andrews2005,Schmidt2010}. After SMF projection, this yields not only signal and pilot coupling efficiencies, but also a relative phase and a mutual-coherence observable. These quantities allow us to test whether the scalar signal-coupling law remains sufficient, or whether a second observable associated with reference tracking is needed to describe the effective key-rate degradation.

The main conclusion of this paper is that the answer depends on the protocol. For GG02, the downstream effective channel is accurately captured by a realistic scalar law for the SMF coupling efficiency within the effective Gaussian closure adopted here, whereas a mean-loss description is generally too optimistic. For the pilot-assisted architecture, the signal coupling law alone is no longer sufficient: signal--pilot phase mismatch and loss of post-coupling coherence generate an additional protocol-dependent reference-noise penalty. In this sense, the present work separates two mechanisms that are often blended together in free-space coherent quantum links: scalar fading of the signal coupling, and differential reference-tracking degradation specific to transmitted-reference architectures.

\subsection{Relation to previous free-space CV-QKD work}
\label{sec:relation-previous-work}

We emphasize that the use of fading-channel statistics, transmittance probability distributions, and moment-matched Gaussian descriptions is not claimed here as a new idea. These tools form an established part of atmospheric and free-space quantum-channel modeling, including CV-QKD over fading links \cite{Vasylyev2012,Usenko2012,Heim2014,Vasylyev2016,Vasylyev2018,Wang2018,Ruppert2019}. The mean-loss-only model used below is therefore not presented as a state-of-the-art free-space CV-QKD model, but as a deliberately simple control case that quantifies the error made when receiver-side fluctuations are ignored.

The specific contribution of the present work is narrower and receiver-focused. We isolate the SMF reception stage by decomposing the effective coupling into aperture transmission and modal compatibility,
\[
\eta_{\mathrm{smf}}=T_{\mathrm{ap}}\Gamma_{\mathrm{mode}},
\]
and we test, with wave-optics Monte Carlo simulations and scalar controls, which part of this optical information remains necessary once the data are passed to the downstream GG02 effective key-rate model. In this sense, the novelty is not that free-space CV-QKD depends on channel statistics, but that the relevant receiver-side statistic is generated by the joint action of aperture collection, modal filtering, and their correlations.

This also clarifies the relation to previous studies of free-space-to-fiber coupling. Quantities closely related to the modal-overlap, fiber-coupling efficiency, or interruption probability have been analyzed before in classical and quantum free-space links \cite{Dikmelik2005,Wang2018,Zuo2020}. Here, that coupling observable is embedded into an effective CV-QKD channel model and compared with controlled alternatives in which the modal factor is frozen, decorrelated, or replaced by scalar surrogate laws. This allows us to distinguish the upstream optical generation of the coupling law from the downstream scalar closure used for GG02.

Finally, regarding transmitted-reference architectures, the delay-dependent model should be read as a sensitivity analysis rather than as a claim that practical signal and pilot pulses always experience independent atmospheric channels. When the signal and the reference are sufficiently close in time and path, the common-channel approximation is appropriate. The present model instead quantifies the residual penalty that appears when this common-mode assumption is imperfect because of a temporal delay, a non-common path, or post-coupling reference mismatch \cite{Qi2015,Soh2015,Marie2017,Laudenbach2019,Ruppert2019}.

The paper is organized as follows. Section~\ref{sec:optical-model} introduces the turbulence-resolved optical propagation model, the SMF reception model, and the signal--pilot extension. Section~\ref{sec:effective-models} defines the effective CV-QKD closures used in the sequel. Section~\ref{sec:numerics} summarizes the numerical methodology and convergence diagnostics. Section~\ref{sec:results-gg02} presents the GG02 results and shows when a scalar coupling law is sufficient. Section~\ref{sec:results-pilot} analyzes the pilot-assisted setting and the emergence of protocol-dependent reference noise. The paper ends with a compact conclusion and outlook in Section~\ref{sec:conclusion}.

\section{Optical model: turbulent propagation and SMF reception}
\label{sec:optical-model}

We consider a free-space optical link in the paraxial regime, with a narrowband field emitted by the transmitter, distorted by atmospheric turbulence, collected by a finite receiver aperture, and finally projected onto a single-mode fiber (SMF) mode. The purpose of this section is to define the optical observables that will later enter the effective CV-QKD channel model. The main point is that, once SMF reception is included, the relevant receiver-side quantity is not the aperture-collected power alone, but an effective coupling observable that combines geometric collection and modal overlap \cite{Andrews2005,Schmidt2010,Dikmelik2005,Vasylyev2016,Vasylyev2018}.

\subsection{Paraxial propagation through turbulence}
\label{sec:paraxial-propagation}

Let \(E(\mathbf{r},z)\), with \(\mathbf{r}=(x,y)\), denote the slowly varying envelope of the optical field at longitudinal coordinate \(z\). In the paraxial approximation, free-space propagation between turbulent perturbations is described by the standard Fresnel propagator \cite{Andrews2005,Schmidt2010}. Numerically, we implement the propagation through a split-step Fourier scheme, alternating free propagation and random phase-screen multiplication, which is a standard approach in wave-optics simulations of atmospheric channels \cite{Andrews2005,Schmidt2010,Vasylyev2012,Vasylyev2016}.

For a total propagation distance \(L\), the path is divided into \(N_{\mathrm{scr}}\) segments of length \(\Delta z=L/N_{\mathrm{scr}}\). At each intermediate plane, the field is multiplied by a random phase factor
\begin{equation}
E(\mathbf{r},z_k^+) = E(\mathbf{r},z_k^-)\exp\!\bigl(i\phi_k(\mathbf{r})\bigr),
\end{equation}
where \(\phi_k\) is generated from a von K\'arm\'an-type turbulence spectrum. Between consecutive screens, the field propagates according to
\begin{equation}
\widehat{E}(\mathbf{q},z+\Delta z)
=
\widehat{E}(\mathbf{q},z)
\exp\!\left(
-\frac{i|\mathbf{q}|^2}{2k}\Delta z
\right),
\end{equation}
with \(k=2\pi/\lambda\) and \(\mathbf{q}=(q_x,q_y)\) the transverse spatial-frequency variable.

The phase screens are parametrized by the refractive-index structure constant \(C_n^2\), together with outer and inner scales \(L_0\) and \(l_0\). Their role here is to provide an effective statistical representation of a turbulent optical path, capturing beam wander, large-scale wavefront distortion, and mode degradation at the receiver \cite{Andrews2005,Schmidt2010,Vasylyev2012,Vasylyev2016}. At the receiver plane \(z=L\), the propagated field is denoted by
\begin{equation}
E_{\mathrm{rec}}(\mathbf{r}) := E(\mathbf{r},L).
\end{equation}

A finite circular aperture of diameter \(D\) selects the collected portion of the beam. If \(A(\mathbf{r})\) denotes the aperture indicator function,
\begin{equation}
A(\mathbf{r})=
\begin{cases}
1, & |\mathbf{r}|\le D/2,\\
0, & |\mathbf{r}|>D/2,
\end{cases}
\end{equation}
the aperture-truncated field is
\begin{equation}
E_A(\mathbf{r}) = A(\mathbf{r})\,E_{\mathrm{rec}}(\mathbf{r}),
\end{equation}
and the aperture-collected power is
\begin{equation}
P_{\mathrm{ap}}=
\int_{\mathbb{R}^2}|E_A(\mathbf{r})|^2\,d^2\mathbf{r}.
\end{equation}

\subsection{Single-mode reception and effective coupling}
\label{sec:smf-coupling}

The receiver is assumed to perform single-mode detection after aperture collection. Let \(u_f(\mathbf{r})\) denote the normalized guided receiving mode in the receiver plane, with
\begin{equation}
\int_{\mathbb{R}^2}|u_f(\mathbf{r})|^2\,d^2\mathbf{r}=1.
\end{equation}
The complex coupling amplitude into the SMF is then
\begin{equation}
a=
\int_{\mathbb{R}^2}E_A(\mathbf{r})\,u_f^*(\mathbf{r})\,d^2\mathbf{r},
\label{eq:coupling-amplitude}
\end{equation}
and the corresponding coupled power is \(|a|^2\).

The effective SMF coupling efficiency is defined as
\begin{equation}
\eta_{\mathrm{smf}}=\frac{|a|^2}{P_{\mathrm{ref}}},
\label{eq:eta-smf-def}
\end{equation}
where \(P_{\mathrm{ref}}\) is the reference power associated with the chosen normalization of the transmitted field. To separate geometric collection from modal degradation, we write
\begin{equation}
\eta_{\mathrm{smf}} = T_{\mathrm{ap}}\,\Gamma_{\mathrm{mode}},
\label{eq:eta-factorization}
\end{equation}
with
\begin{equation}
T_{\mathrm{ap}}=\frac{P_{\mathrm{ap}}}{P_0},
\qquad
\Gamma_{\mathrm{mode}}=\frac{|a|^2}{P_{\mathrm{ap}}},
\end{equation}
where \(P_0\) is the reference received power in the absence of aperture truncation and modal filtering.

The factor \(T_{\mathrm{ap}}\) quantifies geometric collection by the finite receiver aperture, whereas \(\Gamma_{\mathrm{mode}}\) quantifies the compatibility of the collected field with the guided receiving mode. This distinction is central in the present work: the receiver is sensitive not only to how much power is collected, but also to how well the distorted field matches the selected coherent mode \cite{Dikmelik2005,Vasylyev2016,Vasylyev2018}.

\begin{definition}[Effective SMF coupling]
The effective SMF coupling efficiency \(\eta_{\mathrm{smf}}\) is the fraction of optical energy coupled into the receiver single-mode channel after aperture truncation and modal filtering. Its factorization
\[
\eta_{\mathrm{smf}}=T_{\mathrm{ap}}\Gamma_{\mathrm{mode}}
\]
separates geometric collection from modal purity.
\end{definition}

Each turbulent realization therefore produces one value of \(T_{\mathrm{ap}}\), one value of \(\Gamma_{\mathrm{mode}}\), and one value of \(\eta_{\mathrm{smf}}\). Their statistics define the receiver-side optical channel used below.

\subsection{Signal--pilot extension and frozen-flow model}
\label{sec:signal-pilot-model}

To analyze transmitted-reference effects, we extend the optical model to a pair of fields: a signal field \(E_s\) and a pilot field \(E_p\). Both are received through the same aperture and projected onto the same SMF mode, yielding the coupled amplitudes
\begin{equation}
a_{\mathrm{s}}=\langle E_s,u_f\rangle,
\qquad
a_{\mathrm{p}}=\langle E_p,u_f\rangle,
\label{eq:signal-pilot-amps}
\end{equation}
where
\begin{equation}
\langle E,u_f\rangle
=
\int_{\mathbb{R}^2}A(\mathbf{r})\,E(\mathbf{r})\,u_f^*(\mathbf{r})\,d^2\mathbf{r}.
\end{equation}
We define the corresponding signal and pilot coupling efficiencies as
\begin{equation}
\eta_s = |a_{\mathrm{s}}|^2,
\qquad
\eta_p = |a_{\mathrm{p}}|^2.
\end{equation}

In the transmitted-reference setting, the relevant observables are no longer limited to the signal coupling alone. We also introduce the relative phase
\begin{equation}
\Delta\phi=\arg(a_{\mathrm{s}}a_{\mathrm{p}}^*),
\label{eq:delta-phi-def}
\end{equation}
and the complex coherence indicator
\begin{equation}
\nu=
\frac{\left|\mathbb{E}[a_{\mathrm{p}}^*a_{\mathrm{s}}]\right|}
{\sqrt{\mathbb{E}[|a_{\mathrm{s}}|^2]\mathbb{E}[|a_{\mathrm{p}}|^2]}}.
\label{eq:nu-def}
\end{equation}
The quantity \(\nu\) measures how well the transmitted pilot remains locked, after propagation and coupling, to the optical mode seen by the signal.

To generate correlated but non-identical signal and pilot paths, we adopt a frozen-flow construction. The turbulent medium is transported transversely with an effective velocity \(\mathbf{v}\), so that a signal--pilot delay \(\Delta t\) translates into a transverse shift of each phase screen,
\begin{equation}
\Delta\mathbf{r}=\mathbf{v}\,\Delta t.
\label{eq:frozen-flow-shift}
\end{equation}
Thus, \(\Delta t=0\) corresponds to identical signal and pilot propagation, while increasing \(\Delta t\) progressively decorrelates the two optical channels \cite{Taylor1938,Andrews2005,Schmidt2010}. We stress that this delay model is not intended to replace the common-channel approximation used when the signal and reference are rapidly interleaved compared with atmospheric fluctuations; rather, it is used here to quantify the sensitivity of a transmitted-reference architecture to residual differential propagation and imperfect common-mode rejection \cite{Ruppert2019}.

This extension provides the minimal optical structure required to distinguish scalar signal fading from differential signal--pilot decoherence. In the GG02 setting, the main receiver-side observable will be the effective signal coupling law. In the transmitted-reference setting, the additional quantities \(\Delta\phi\) and \(\nu\) will serve as optical indicators of reference degradation.

\section{Effective channel models for CV-QKD}
\label{sec:effective-models}

The optical model of Section~\ref{sec:optical-model} produces, for each turbulent realization, receiver-side observables defined after aperture truncation and single-mode projection. The purpose of the present section is to convert these optical quantities into effective parameters entering a CV-QKD key-rate model. Our aim is not to provide a complete security treatment of every architecture considered below, but rather to define a hierarchy of effective closures that makes it possible to identify which optical information is required for key-rate prediction in the present setting \cite{Grosshans2002,Grosshans2003,GarciaPatron2006,Leverrier2013,Leverrier2015,Leverrier2017,Laudenbach2018,Pirandola2020}.

Throughout the paper, we work in an effective asymptotic reverse-reconciliation setting with reconciliation efficiency \(\beta\). The GG02 protocol is used as the baseline Gaussian-modulated coherent-state reference model \cite{Grosshans2002,Grosshans2003,GarciaPatron2006,Leverrier2013,Leverrier2015,Laudenbach2018}. In the GG02 setting, the relevant receiver-side random variable is the effective signal coupling efficiency \(\eta_{\mathrm{smf}}\). In the pilot-assisted extension, the signal channel remains governed by the same type of observable, but an additional reference-noise contribution appears through the signal-pilot differential quantities introduced in Section~\ref{sec:signal-pilot-model} \cite{Qi2015,Soh2015,Marie2017,Laudenbach2019}.

\subsection{Baseline GG02 effective models}
\label{sec:gg02-effective-models}

We first consider an effective Gaussian channel described by a transmittance \(T\) and an excess noise \(\xi\), both referred to the channel input. Let
\begin{equation}
V = V_A + 1
\end{equation}
be Alice's total quadrature variance in shot-noise units, where \(V_A\) is the modulation variance. In the equivalent entanglement-based description, we use the standard Gaussian covariance matrix
\begin{equation}
\gamma_{AB}(T,\xi)=
\begin{pmatrix}
V\,\mathbb{I}_2 & \sqrt{T}\,Z\,\sigma_z \\
\sqrt{T}\,Z\,\sigma_z & \bigl(T(V-1+\xi)+1\bigr)\mathbb{I}_2
\end{pmatrix},
\label{eq:effective-covariance}
\end{equation}
with
\begin{equation}
Z=\sqrt{V^2-1},
\qquad
\mathbb{I}_2=
\begin{pmatrix}
1&0\\
0&1
\end{pmatrix},
\qquad
\sigma_z=
\begin{pmatrix}
1&0\\
0&-1
\end{pmatrix}.
\end{equation}
This is the standard effective Gaussian description used in coherent-state CV-QKD \cite{Grosshans2002,Grosshans2003,GarciaPatron2006,Leverrier2013,Leverrier2015,Pirandola2020}.

The corresponding secret key rate is written as
\begin{equation}
K=\beta I_{AB}-\chi_{BE},
\label{eq:skr-definition}
\end{equation}
where \(I_{AB}\) is the classical mutual information between Alice and Bob and \(\chi_{BE}\) is the Holevo information between Bob and Eve computed from the symplectic spectrum of \(\gamma_{AB}(T,\xi)\) in the usual Gaussian formalism \cite{GarciaPatron2006,Leverrier2013,Leverrier2015,Leverrier2017,Laudenbach2018}. In the present work, this downstream formula is used as an effective closure whose inputs are inferred from the optical simulations.

Let \(\eta\) denote the random effective signal coupling, identified with \(\eta_{\mathrm{smf}}\) in the GG02 setting. We consider three levels of effective closure.

\paragraph{Mean-loss-only closure.}
The simplest approximation replaces the fluctuating channel by a deterministic attenuator with
\begin{equation}
T_{\mathrm{mean}}=\mathbb{E}[\eta],
\qquad
\xi_{\mathrm{mean}}=0.
\label{eq:mean-loss-closure}
\end{equation}
This model discards all coupling fluctuations and keeps only the average transmission. Such approximations are useful as a first baseline, but they do not account for fluctuation-induced excess noise in fading channels \cite{Vasylyev2016,Vasylyev2018,Chai2019,Hosseinidehaj2021}.
In the present work, this closure is used only as a diagnostic lower-complexity baseline. It is not intended to replace the standard fading-channel treatment based on the statistics of the channel transmittance or on moment-matched effective Gaussian parameters.

\paragraph{Mean-\(T\) fading closure.}
A first fluctuation-aware model keeps the average transmittance
\begin{equation}
T_{\mathrm{fading}}=\mathbb{E}[\eta],
\end{equation}
but adds an excess-noise contribution associated with amplitude fluctuations:
\begin{equation}
\xi_{\mathrm{fading}}^{(\mathrm{mean}T)}
=
V_A\,
\frac{\mathrm{Var}(\sqrt{\eta})}{\mathbb{E}[\eta]}.
\label{eq:xi-fading-meanT}
\end{equation}
Equivalently,
\begin{equation}
\xi_{\mathrm{fading}}^{(\mathrm{mean}T)}
=
V_A\,
\frac{\mathbb{E}[\eta]-\mathbb{E}[\sqrt{\eta}]^2}{\mathbb{E}[\eta]}.
\end{equation}
This closure reflects the fact that coherent detection is sensitive to field amplitudes and therefore to \(\sqrt{\eta}\), rather than to \(\eta\) alone \cite{Vasylyev2016,Vasylyev2018,Chai2019}.

\paragraph{Moment closure.}
A more faithful Gaussian effective model is obtained by matching the leading amplitude moments. We define
\begin{equation}
T_{\mathrm{mom}}=\mathbb{E}[\sqrt{\eta}]^2,
\label{eq:T-moment}
\end{equation}
and
\begin{equation}
\xi_{\mathrm{fading}}^{(\mathrm{mom})}
=
V_A\,
\frac{\mathbb{E}[\eta]-\mathbb{E}[\sqrt{\eta}]^2}{\mathbb{E}[\sqrt{\eta}]^2}
=
V_A\,
\frac{\mathrm{Var}(\sqrt{\eta})}{T_{\mathrm{mom}}}.
\label{eq:xi-moment}
\end{equation}
This closure preserves the effective coherent amplitude transfer more faithfully than the mean-loss-only model and will serve as the main scalar baseline in the results below \cite{Vasylyev2018,Chai2019,Hosseinidehaj2021}.

These three closures form a simple hierarchy:
\begin{itemize}
    \item the mean-loss-only model ignores fluctuations altogether,
    \item the mean-\(T\) fading model keeps \(T=\mathbb{E}[\eta]\) and adds fluctuation-induced noise,
    \item the moment closure uses the amplitude-compatible effective transmittance \(T_{\mathrm{mom}}=\mathbb{E}[\sqrt{\eta}]^2\).
\end{itemize}
The results below will show that the first level is generally too optimistic, while the third provides a much more accurate effective description.

\subsection{Scalar surrogates for the coupling law}
\label{sec:scalar-surrogates}

The previous subsection explains how a given coupling random variable \(\eta\) is converted into an effective GG02 channel. A separate question is whether the full optical propagation problem can itself be reduced to a scalar law for \(\eta\). This is the role of the scalar surrogates considered here.

Let \(\eta_{\mathrm{full}}\) denote the coupling samples generated by the complete turbulent propagation and SMF reception model. We compare this full optical output with several scalar surrogate laws constructed from the same data:
\begin{enumerate}
    \item an empirical surrogate obtained by resampling the simulated values of \(\eta_{\mathrm{full}}\),
    \item a Beta surrogate adjusted to the first two moments of \(\eta_{\mathrm{full}}\),
    \item a truncated lognormal surrogate,
    \item a truncated normal surrogate.
\end{enumerate}
These surrogates are introduced to test whether the downstream GG02 model depends essentially on the scalar statistics of the effective coupling, or whether it retains traces of a more structured optical dependence \cite{Vasylyev2016,Vasylyev2018,Chai2019}.

If a surrogate law yields the same effective key rate as the full optical channel, then the corresponding GG02 effective channel may legitimately be regarded as scalar at the downstream level, even though the scalar law itself was generated by a nontrivial propagation and modal-filtering mechanism. This distinction between upstream optical generation and downstream scalar closure will be central in the interpretation of the GG02 results.

\subsection{Pilot-assisted effective reference-noise model}
\label{sec:pilot-assisted-effective-model}

We now turn to the transmitted-reference setting introduced in Section~\ref{sec:signal-pilot-model}. In this case, the signal still experiences a fluctuating coupling efficiency \(\eta_s\), so that the fading part of the effective channel is treated exactly as in the GG02 moment closure:
\begin{equation}
T_{\mathrm{sig}}=\mathbb{E}[\sqrt{\eta_s}]^2,
\qquad
\xi_{\mathrm{fading}}=
V_A\,
\frac{\mathrm{Var}(\sqrt{\eta_s})}{\mathbb{E}[\sqrt{\eta_s}]^2}.
\label{eq:signal-fading-pilot}
\end{equation}
Thus, the signal branch is kept as close as possible to the scalar GG02 baseline.

The protocol dependence enters because the reference is no longer assumed to track the signal perfectly. In addition to scalar fading, the receiver experiences a differential signal-pilot degradation which we convert into an effective reference-noise contribution \(\xi_{\mathrm{ref}}\). We therefore write
\begin{equation}
\xi_{\mathrm{tot}}=\xi_{\mathrm{fading}}+\xi_{\mathrm{ref}}.
\label{eq:xi-total}
\end{equation}
This decomposition separates the scalar signal-channel contribution from the protocol-dependent reference contribution \cite{Qi2015,Soh2015,Marie2017,Laudenbach2019}.

We consider two effective closures for \(\xi_{\mathrm{ref}}\).

\paragraph{Phase-variance closure.}
Using the relative phase
\begin{equation}
\Delta\phi=\arg(a_{\mathrm{s}}a_{\mathrm{p}}^*),
\end{equation}
we define
\begin{equation}
\xi_{\mathrm{ref}}^{(\phi)}
=
V_A\,\mathrm{Var}(\Delta\phi).
\label{eq:xi-ref-phase}
\end{equation}
In practice, this is implemented through an effective wrapped-phase variance when the phase distribution remains sufficiently concentrated. This closure emphasizes the phase-tracking interpretation of the pilot \cite{Qi2015,Soh2015,Marie2017}.
This expression follows from the usual small phase-error approximation. If the pilot defines the phase reference used by Bob, a residual error \(\delta\phi\) rotates the measured quadrature according to \(X\mapsto X\cos\delta\phi+P\sin\delta\phi\). For \(|\delta\phi|\ll 1\), this gives an additive quadrature perturbation \(\delta X\simeq \delta\phi P\). When the signal modulation scale is \(V_A\), the corresponding effective input-referred excess noise is proportional to \(V_A\mathrm{Var}(\delta\phi)\), which leads to Eq.~\eqref{eq:xi-ref-phase} within the convention used here.

\paragraph{Coherence closure.}
Using the complex coherence indicator
\begin{equation}
\nu=
\frac{\left|\mathbb{E}[a_{\mathrm{p}}^*a_{\mathrm{s}}]\right|}
{\sqrt{\mathbb{E}[|a_{\mathrm{s}}|^2]\mathbb{E}[|a_{\mathrm{p}}|^2]}},
\end{equation}
we define
\begin{equation}
\xi_{\mathrm{ref}}^{(\nu)}
=
2V_A(1-\nu).
\label{eq:xi-ref-nu}
\end{equation}
This closure emphasizes the loss of coherent compatibility between signal and pilot after propagation and single-mode projection \cite{Marie2017,Laudenbach2019}.
The factor \(2(1-\nu)\) can be obtained from a normalized differential-field error. Let
\[
\tilde a_s=\frac{a_s}{\sqrt{\mathbb{E}[|a_s|^2]}},
\qquad
\tilde a_p=\frac{a_p}{\sqrt{\mathbb{E}[|a_p|^2]}},
\]
and choose the constant phase of the pilot so as to maximize the real overlap with the signal. The residual mean-square mismatch is then
\begin{align}
\mathcal{E}_{\nu}
&=\min_{\theta}\,\mathbb{E}\!\left[\left|\tilde a_s-e^{i\theta}\tilde a_p\right|^2\right]\\
&=2-2\left|\mathbb{E}[\tilde a_p^*\tilde a_s]\right|\\
&=2(1-\nu).
\end{align}
Interpreting this residual field mismatch as a reference-induced quadrature error on the modulation scale \(V_A\) gives \(\xi_{\mathrm{ref}}^{(\nu)}=V_A\mathcal{E}_{\nu}=2V_A(1-\nu)\). This derivation is now included to make explicit that Eq.~\eqref{eq:xi-ref-nu} is an effective coherence-penalty model, not an additional fundamental security theorem.

The corresponding pilot-assisted effective rates are obtained by inserting
\begin{equation}
(T,\xi)=
\bigl(T_{\mathrm{sig}},\,\xi_{\mathrm{fading}}+\xi_{\mathrm{ref}}^{(\phi)}\bigr)
\end{equation}
or
\begin{equation}
(T,\xi)=
\bigl(T_{\mathrm{sig}},\,\xi_{\mathrm{fading}}+\xi_{\mathrm{ref}}^{(\nu)}\bigr)
\end{equation}
into the same downstream Gaussian key-rate model, Eqs.~\eqref{eq:effective-covariance} and \eqref{eq:skr-definition}. In this way, the signal channel itself is kept identical to the scalar baseline, while the protocol dependence is isolated in the additional reference-noise term.

\begin{remark}
The pilot-assisted closures in Eqs.~\eqref{eq:xi-ref-phase} and \eqref{eq:xi-ref-nu} are effective physical models intended to isolate the role of signal-pilot differential decoherence. They are not a complete composable security treatment of a realistic transmitted-reference protocol.
\end{remark}

This construction separates two physically distinct effects:
\begin{enumerate}
    \item a scalar fading contribution, already present in the GG02 baseline and entirely determined by the signal coupling law,
    \item a protocol-dependent reference contribution, controlled by the differential observables \(\Delta\phi\) and \(\nu\).
\end{enumerate}
This separation will allow us to identify below when a scalar description remains sufficient and when an additional differential reference observable becomes necessary.

\section{Numerical methodology and convergence diagnostics}
\label{sec:numerics}

This section summarizes the numerical procedure used to generate the optical observables and the effective key-rate estimates reported below. The purpose is not to provide a full numerical analysis of all discretization issues, but to state the computational choices that matter for the physical interpretation of the results. Since the effective CV-QKD parameters are extracted from Monte Carlo samples of turbulence-resolved optical propagation, convergence with respect to the number of realizations is an essential part of the modeling procedure \cite{Andrews2005,Schmidt2010,Vasylyev2012,Vasylyev2016,Vasylyev2018}.

\begin{table*}[t]
\begingroup\small
\caption{Main numerical parameters of the wave-optics and CV-QKD simulations. The turbulence scans use a fixed optical geometry while varying \(C_n^2\). The delay scan is performed at \(C_n^2=3.0\times 10^{-15}\,\mathrm{m}^{-2/3}\) with five independent seeds and uncertainty estimates.}
\label{tab:numerical-parameters}
\resizebox{\textwidth}{!}{%
\begin{tabular}{@{}lll@{}}
\hline
Parameter & Symbol & Value used in the simulations \\
\hline
Propagation distance & \(L\) & \(1000\,\mathrm{m}\) \\
Optical wavelength & \(\lambda\) & \(1550\,\mathrm{nm}\) \\
Input beam waist & \(w_0\) & \(20\,\mathrm{mm}\) \\
Receiver aperture diameter & \(D\) & \(100\,\mathrm{mm}\) \\
SMF mode waist at receiver & \(w_f\) & \(20\,\mathrm{mm}\) \\
Transverse grid size & \(N\times N\) & \(256\times256\) \\
Computational window & \(L_x\times L_x\) & \(0.30\,\mathrm{m}\times0.30\,\mathrm{m}\) \\
Grid spacing & \(\Delta x\) & \(1.17\,\mathrm{mm}\) \\
Number of phase screens & \(N_{\mathrm{scr}}\) & 5, midpoint split-step propagation \\
Turbulence strengths & \(C_n^2\) & \(0,5\times10^{-16},10^{-15},3\times10^{-15},5\times10^{-15}\,\mathrm{m}^{-2/3}\) \\
Outer and inner scales & \(L_0,l_0\) & \(30\,\mathrm{m}\), \(5\,\mathrm{mm}\) \\
Subharmonic correction & -- & included, 3 subharmonic levels \\
Monte Carlo realizations & \(N_{\mathrm{MC}}\) & 300 per turbulence value \\
Delay-scan statistics & -- & 300 realizations per delay and per seed, 5 seeds \\
Frozen-flow velocity & \(v\) & \(10\,\mathrm{m\,s^{-1}}\), along the \(x\)-axis \\
Frozen-flow shift & -- & Fourier sub-pixel shift in the delay scan \\
Signal-pilot delays & \(\Delta t\) & \(0,2\times10^{-5},5\times10^{-5},8\times10^{-5},10^{-4},1.2\times10^{-4},1.5\times10^{-4},\)\newline\(2\times10^{-4},3\times10^{-4},5\times10^{-4},10^{-3}\,\mathrm{s}\) \\
Modulation variance & \(V_A\) & 4 SNU \\
Reconciliation efficiency & \(\beta\) & 0.95 \\
Additional baseline technical excess noise & \(\xi_0\) & not included in the reported effective-rate comparisons \\
\hline
\end{tabular}%
}
\endgroup
\end{table*}

\subsection{Transverse grid and propagation settings}
\label{sec:grid-and-parameters}

The propagated field is represented on a square transverse grid of size \(N\times N\), with a finite computational window chosen large enough to accommodate diffraction broadening, beam wander, and aperture truncation over the propagation distance of interest. The numerical values used in the simulations are selected so as to resolve simultaneously the receiver aperture, the guided receiving mode, and the dominant turbulent distortions generated by the phase screens. This compromise between spatial resolution, propagation accuracy, and computational cost is standard in split-step simulations of atmospheric optical propagation \cite{Andrews2005,Schmidt2010}.

The optical wavelength, beam waist, receiver aperture diameter, and receiving-mode waist are fixed throughout a given simulation campaign. The propagation path is divided into a finite number of phase-screen steps, and free-space propagation between screens is implemented through Fourier-domain Fresnel transfer. In the results reported here, the propagation distance, aperture geometry, and number of phase screens are held fixed within each turbulence scan, so that the effect of turbulence strength can be isolated from purely geometric changes in the optical link \cite{Andrews2005,Schmidt2010,Vasylyev2012}.

\subsection{Phase-screen generation}
\label{sec:phase-screen-generation}

Each turbulent realization is generated from a sequence of random phase screens based on a von K\'arm\'an-type spectrum with prescribed refractive-index structure constant \(C_n^2\), outer scale, and inner scale. This is a standard choice for wave-optics simulations of atmospheric channels, since it captures both inertial-range distortions and the finite-scale cutoffs relevant for realistic beam propagation \cite{Andrews2005,Schmidt2010}.
To improve the representation of low-spatial-frequency distortions, especially beam wander and large-scale wavefront tilts, the FFT-based screen synthesis is supplemented by subharmonic corrections.
Such low-frequency enhancements are well known to improve the realism of finite-grid turbulence simulations when receiver performance is sensitive to centroid motion and large-scale phase distortions \cite{Lane1992,Andrews2005,Schmidt2010}.

The propagation itself is implemented through a split-step Fourier scheme. In the numerical version used here, each longitudinal step is written in midpoint form, with a half-step of free propagation on each side of the phase-screen multiplication. This improves the symmetry of the propagation and reduces bias relative to a purely one-sided scheme. Although the present paper is not focused on numerical analysis per se, this construction provides a reliable and efficient propagation model in the turbulence regime considered here \cite{Andrews2005,Schmidt2010,Vasylyev2012}.

\subsection{Receiver-side observables}
\label{sec:receiver-observables}

For each turbulent realization, the propagated field is truncated by the receiver aperture and then projected onto the receiving SMF mode. This yields the aperture transmission factor \(T_{\mathrm{ap}}\), the modal-purity factor \(\Gamma_{\mathrm{mode}}\), and the effective coupling efficiency \(\eta_{\mathrm{smf}}\). These are the primary optical outputs of the simulation in the GG02 setting. Their roles are distinct: \(T_{\mathrm{ap}}\) measures geometric collection, while \(\Gamma_{\mathrm{mode}}\) measures the compatibility of the collected field with the selected coherent mode. Their product defines the actual scalar coupling variable used in the downstream effective channel model \cite{Dikmelik2005,Vasylyev2016,Vasylyev2018}.

In the pilot-assisted extension, the same receiver-side projection is applied separately to the signal and pilot fields. This yields the coupled complex amplitudes of the signal and pilot, from which we compute not only the corresponding coupling efficiencies but also the relative phase and coherence observables entering the effective reference-noise model. Thus, the same optical propagation engine underlies both the scalar GG02 analysis and the protocol-dependent transmitted-reference extension \cite{Qi2015,Soh2015,Marie2017,Laudenbach2019}.

\subsection{Frozen-flow implementation}
\label{sec:frozen-flow-implementation}

In the pilot-assisted simulations, the signal and pilot are propagated through correlated but non-identical turbulence realizations generated by a frozen-flow construction. Operationally, each pilot phase screen is obtained by shifting the corresponding signal screen by an amount proportional to the effective wind velocity and to the signal-pilot delay. This is a direct numerical implementation of the standard frozen-flow viewpoint, in which temporal decorrelation is modeled through transverse advection of turbulence patterns \cite{Taylor1938,Andrews2005,Schmidt2010}.
For the delay scan, the transverse shift is implemented as a Fourier sub-pixel shift rather than as an integer-pixel roll. This removes the artificial discontinuity associated with rounding the frozen-flow displacement to the nearest grid point. The scan is also repeated over five independent random seeds, and the figures report the mean value with the standard error across seeds.

This construction has two practical advantages in the present context. First, it introduces a physically transparent control parameter, namely the signal-pilot delay. Second, it allows one to vary the degree of signal-pilot mismatch without changing the underlying turbulence strength or the receiver geometry. As a result, the protocol-dependent degradation observed in the pilot-assisted key rates can be associated directly with differential reference tracking rather than with changes in the scalar signal channel itself \cite{Qi2015,Soh2015,Marie2017,Laudenbach2019}.

\subsection{Monte Carlo estimation of effective channel parameters}
\label{sec:monte-carlo-estimation}

All effective channel quantities are estimated from ensembles of independent turbulent realizations. In the GG02 setting, the relevant sample moments are the mean and variance of the coupling efficiency and, more importantly, the moments of its square root, since the coherent amplitude transfer is governed by \(\sqrt{\eta_{\mathrm{smf}}}\) rather than by \(\eta_{\mathrm{smf}}\) alone \cite{Vasylyev2016,Vasylyev2018,Chai2019}. These sample moments determine the effective transmittance and excess-noise parameters used in the downstream Gaussian key-rate model.

In the pilot-assisted setting, additional ensemble averages are required, including the mean pilot coupling, the wrapped relative-phase statistics, and the complex coherence between the signal and pilot coupled amplitudes. These quantities are then converted into effective reference-noise terms according to the closures introduced in Section~\ref{sec:pilot-assisted-effective-model}. The numerical procedure is therefore modular: the optical Monte Carlo stage produces a statistically rich set of receiver-side observables, and the effective CV-QKD stage translates them into key-rate estimates \cite{Qi2015,Soh2015,Marie2017,Laudenbach2019,Vasylyev2018}.

\subsection{Convergence diagnostics}
\label{sec:convergence-diagnostics}

Because the effective key rate depends nonlinearly on empirical channel moments, convergence of the Monte Carlo estimates must be checked explicitly. In particular, one should distinguish between the convergence of low-order optical moments, such as \(\mathbb{E}[\eta_{\mathrm{smf}}]\), and the convergence of the final key rate. The former is typically faster than the latter, especially near threshold, where small errors in effective excess noise may produce amplified changes in the key rate \cite{Leverrier2013,Leverrier2015,Vasylyev2018,Hosseinidehaj2021}.

For this reason, we monitor the cumulative behavior of representative observables as the number of realizations increases. These diagnostics include the mean coupling, the coupling variance, the effective excess-noise terms, and the full key-rate estimates in representative turbulence regimes. The convergence behavior follows the expected hierarchy: the mean coupling stabilizes first, then the variance-related observables, and finally the key rate itself. A representative convergence plot is provided in Appendix~\ref{app:convergence}.

In practice, the sample sizes used in the main text were chosen so that the key-rate curves were stable at the scale visible in the reported figures. The purpose of the convergence analysis is therefore not to claim machine-precision estimation, but to ensure that the comparisons made in the main text, including mean loss versus scalar fading and scalar baseline versus pilot-assisted models, are not artifacts of insufficient sampling \cite{Vasylyev2018,Chai2019,Hosseinidehaj2021}.

\subsection{Scope of the numerical model}
\label{sec:scope-numerical-model}

The simulations used in this work are intended as a physically informed effective-modeling framework rather than as a full end-to-end description of a specific deployed system. The optical part of the model resolves turbulent propagation, finite-aperture collection, and single-mode projection, while the QKD part uses asymptotic Gaussian effective closures for the signal channel and for the pilot-assisted reference-noise contribution. This level of modeling is sufficient for the purpose of the present paper, namely identifying which optical observables control the effective channel in GG02 and when an additional protocol-dependent coherence observable becomes necessary in transmitted-reference architectures \cite{Grosshans2002,Leverrier2013,Leverrier2015,Vasylyev2018,Qi2015,Soh2015,Marie2017,Laudenbach2019}.

The model should not be read as an experimental validation of a particular deployed free-space CV-QKD link. Several important impairments are not explicitly optimized here, including pointing and tracking errors, absorption, scattering, background light, detector imperfections, adaptive-optics control bandwidth, and full link-budget margins. These effects can be included at the effective level as additional loss or excess-noise contributions, or at the optical level through more detailed propagation and receiver models. The present simulations isolate one specific mechanism: turbulence-induced wavefront distortion followed by finite-aperture collection and SMF modal filtering. This isolation is useful because it identifies which receiver-side optical observables should be carried into the effective CV-QKD model before other engineering impairments are added.

With these numerical ingredients in place, we now turn to the main physical results. We first analyze the GG02 baseline and test whether the full optical propagation can be reduced to a realistic scalar law for the effective coupling. We then examine how this picture changes when a transmitted reference is introduced.

\section{Results I: effective coupling statistics for GG02}
\label{sec:results-gg02}

We first consider the GG02 baseline and ask a simple reduction question: once turbulent propagation, finite-aperture collection, and single-mode reception are modeled explicitly, how much optical information must be retained in order to reproduce the effective secret key rate? This issue is central in free-space CV-QKD because atmospheric channels are often described through average transmittance or coarse fading models, whereas coherent single-mode reception naturally produces a receiver-side observable with a more structured optical origin \cite{Heim2014,Vasylyev2016,Vasylyev2018,Chai2019,Hosseinidehaj2021,Pirandola2021}.

The results reported in this section show a clear hierarchy. A mean-loss-only description is generally too optimistic. By contrast, a realistic scalar law for the effective SMF coupling reproduces the downstream GG02 channel with good accuracy over a broad turbulence range. At the same time, the optical factorization of the coupling into aperture transmission and modal purity remains physically important, because it explains how that scalar law is generated by the propagation problem \cite{Dikmelik2005,Vasylyev2016,Vasylyev2018}.

\subsection{Mean loss alone is insufficient}
\label{sec:mean-loss-insufficient}

We begin by comparing the full optical model with the simplest effective approximation, namely a deterministic channel obtained by replacing the fluctuating coupling by its average value. This approximation systematically overestimates the secret key rate, and the discrepancy becomes increasingly pronounced as the turbulence level grows. This behavior is consistent with the general understanding that atmospheric fading affects CV-QKD not only through average transmission, but also through fluctuation-induced excess noise and degradation of coherent amplitude transfer \cite{Vasylyev2016,Vasylyev2018,Chai2019,Hosseinidehaj2021}.

The reason is straightforward. The mean-loss-only model retains the average coupled energy but discards the fluctuations of the effective field amplitude, which are precisely the quantities entering the excess-noise terms of the Gaussian downstream closure. Since coherent detection is sensitive to amplitude transfer rather than to power alone, a channel with the correct mean coupling but no coupling fluctuations is generally too favorable. This is the optical origin of the failure of mean-loss descriptions in the present setting.

This effect is already visible in the intermediate turbulence regime and becomes severe in the strongest cases considered here. In practice, the mean-loss-only model can remain comfortably positive even when the full fluctuating model is already close to threshold. From the viewpoint of effective modeling, this establishes a basic point: the relevant channel variable is not the average attenuation \(\mathbb{E}[\eta_{\mathrm{smf}}]\) alone, but the full statistics of the receiver-side coupling observable generated by turbulent propagation and modal filtering.

\subsection{A scalar coupling law is sufficient for GG02}
\label{sec:scalar-law-sufficient}

We now turn to the main GG02 result. Although mean loss alone is insufficient, the downstream GG02 key-rate model does not appear to require an intrinsically non-scalar channel description once the correct coupling statistics have been extracted from the optical simulations. More precisely, the full optical model is accurately reproduced by scalar surrogate laws for the effective SMF coupling efficiency, provided these surrogates capture the relevant shape of the coupling distribution.

This statement is supported by the comparison between the full optical simulations and several scalar surrogates: empirical resampling of the simulated coupling values, a Beta fit adjusted to the first two moments, and truncated lognormal or normal surrogates. Across weak and moderate turbulence, all of these scalar descriptions remain close to the full model, and the empirical surrogate is nearly indistinguishable from it. This indicates that the downstream GG02 Gaussian closure reacts primarily to the scalar statistics of the effective coupling, not to a more microscopic memory of the propagation path itself \cite{Vasylyev2016,Vasylyev2018,Chai2019,Hosseinidehaj2021}.

The agreement is especially clear in the intermediate regime, where the full model and the empirical scalar surrogate yield essentially the same effective key rate. Even in the strongest turbulence case considered here, the various scalar surrogates still reproduce the same qualitative behavior, namely a strong collapse of the key rate, whereas the mean-loss-only model remains far too optimistic. The remaining discrepancies between scalar surrogates in this regime appear to be associated mainly with the detailed shape of the coupling tails.

\begin{figure}[htbp]
    \centering
    \includegraphics[width=.4\textwidth]{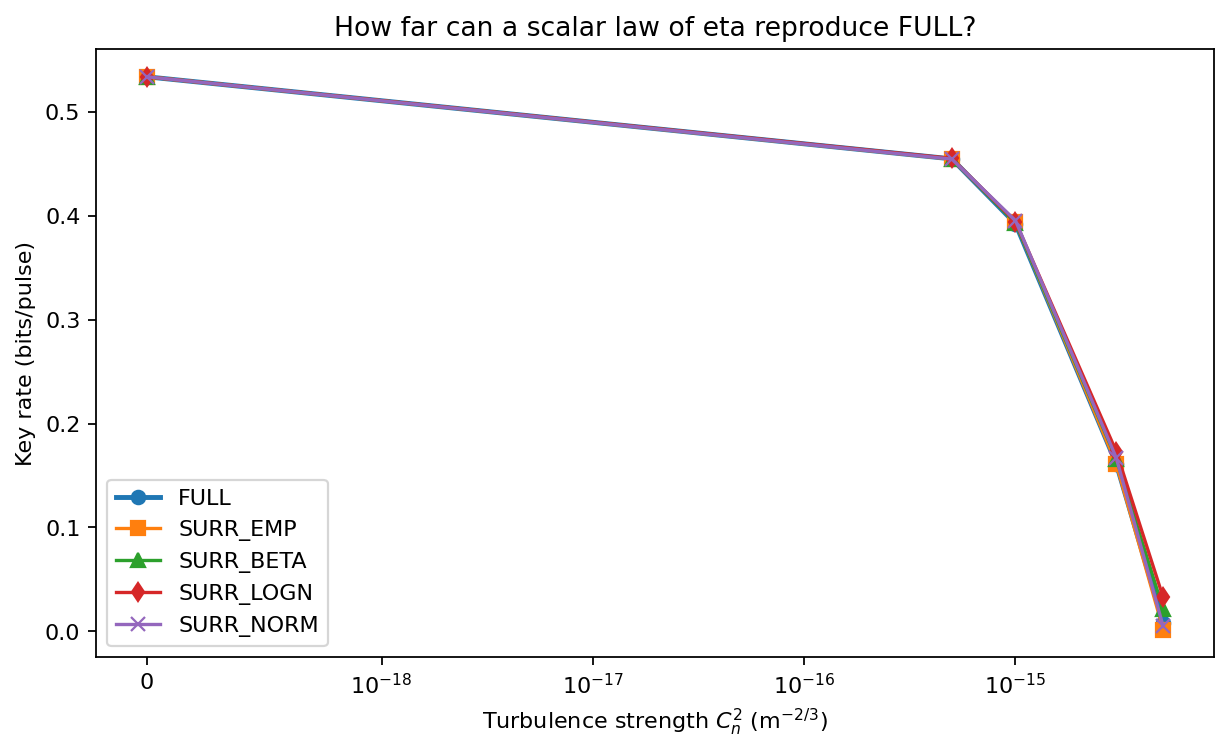}
    \caption{Comparison between the full optical model, a mean-loss-only approximation, and several scalar surrogates for the effective SMF coupling law in the GG02 setting. Mean loss alone systematically overestimates the secret key rate, whereas realistic scalar coupling laws reproduce the full model with good accuracy over a broad turbulence range.}
    \label{fig:scalar-keyrates}
\end{figure}

The conclusion is therefore precise. The effective GG02 channel is not well described by average attenuation alone, but once a realistic scalar law for the receiver-side coupling has been obtained from the optical model, this law is largely sufficient for effective GG02 key-rate prediction. The full propagation model remains essential upstream because it generates the physically correct coupling statistics under turbulence. Downstream, however, the GG02 model largely reads those statistics through a scalar effective channel. This behavior is illustrated in Fig.~\ref{fig:scalar-keyrates} \cite{Grosshans2002,Grosshans2003,Leverrier2013,Leverrier2015,Vasylyev2018,Chai2019}.

This point is conceptually important. The present simulations do not support the stronger claim that the GG02 effective channel remains irreducibly non-scalar after turbulent propagation and SMF reception. Rather, they support a more precise statement: the physically relevant scalar variable is not the mean attenuation, but the effective coupling law produced by the full optical propagation and modal-filtering mechanism.

\subsection{Optical origin of the scalar coupling law}
\label{sec:factorization-controls}

The fact that a scalar law for the effective coupling is sufficient for GG02 does not mean that the optical factorization
\[
\eta_{\mathrm{smf}} = T_{\mathrm{ap}}\,\Gamma_{\mathrm{mode}}
\]
is physically irrelevant. On the contrary, the control studies show that the structure of this factorization matters for the generation of the scalar coupling law itself. This is where the explicit optical model carries information that would be lost in a purely phenomenological fading parametrization \cite{Dikmelik2005,Vasylyev2016,Vasylyev2018}.

The first observation is that the modal-purity factor \(\Gamma_{\mathrm{mode}}\) is not a trivial function of the aperture transmission \(T_{\mathrm{ap}}\). As turbulence increases, both quantities degrade, but their fluctuations are neither independent nor perfectly locked. The corresponding scatter plots show a clear positive trend together with a substantial residual dispersion. Thus, the modal purity cannot be reduced to a deterministic function of the aperture-collected power. This is physically consistent with the distinction between geometric collection losses and turbulence-induced mode mismatch in free-space-to-fiber coupling \cite{Dikmelik2005,Andrews2005,Schmidt2010}.

To quantify this point, we introduced control models in which the modal-purity factor is either frozen at its empirical mean or decorrelated from the aperture-transmission factor while preserving the marginal fluctuations. Both controls yield systematically more optimistic key-rate estimates than the full model, especially in the moderate and strong turbulence regimes. This shows that two ingredients matter in practice. First, the fluctuations of the modal-purity factor contribute non-negligibly to the effective coupling statistics. Second, the positive correlation between aperture transmission and modal purity further enhances the degradation of the coupled channel.

The correct interpretation of the GG02 results is therefore the following. The full optical propagation problem generates a nontrivial scalar law for the effective coupling through the joint action of aperture transmission, modal purity, and their correlation. Once this law has been constructed, the downstream GG02 effective channel can be described to good accuracy by a scalar surrogate. Thus, the optical factorization remains physically important upstream, even though the final GG02 description is largely scalar downstream. Additional control results are reported in Appendix~\ref{app:additional-controls}.

Taken together, the results of this section identify the appropriate level of effective description for GG02 with SMF reception. Mean loss alone is insufficient. A realistic scalar law for the effective coupling is sufficient. And the physical origin of that law lies in the coupled action of geometric collection and modal filtering under atmospheric turbulence \cite{Dikmelik2005,Vasylyev2016,Vasylyev2018,Chai2019,Hosseinidehaj2021}.

\section{Results II: protocol dependence and transmitted-reference noise}
\label{sec:results-pilot}

We now turn to the pilot-assisted setting and ask whether the scalar signal-coupling description identified in the GG02 analysis remains sufficient once the receiver relies on a transmitted reference. This question is directly relevant to coherent CV-QKD architectures in which reference recovery is part of the optical communication problem itself \cite{Qi2015,Soh2015,Marie2017,Laudenbach2019,Pirandola2020}. The results reported below show that the answer is no: in a pilot-assisted architecture, the atmosphere does not merely generate scalar signal fading, but also a differential signal-pilot degradation that appears as an additional protocol-dependent reference-noise penalty.

\subsection{Comparison at fixed signal-pilot delay}
\label{sec:fixed-delay-results}

We first compare the scalar GG02 baseline with the pilot-assisted effective models at a fixed signal-pilot delay. In this configuration, the signal and pilot propagate through correlated but distinct turbulent realizations generated by the frozen-flow construction introduced in Section~\ref{sec:signal-pilot-model}. The delay is chosen so that the signal and pilot remain strongly correlated, while still exhibiting a measurable differential mismatch.

A first observation is that the pilot-assisted curves lie systematically below the scalar baseline. This effect is weak in the low-turbulence regime, where the pilot still follows the signal closely, but it becomes progressively more pronounced as the turbulence level increases and the system approaches threshold. Thus, once the protocol depends on a transmitted reference, the scalar signal-coupling law is no longer the only relevant optical input for key-rate estimation.

This behavior has a direct physical interpretation. The scalar baseline only sees the signal channel through the effective law of the signal coupling. By contrast, the pilot-assisted models also include a differential signal-pilot contribution through either the relative-phase fluctuations or the coherence loss. These additional terms act as protocol-dependent excess noise, thereby degrading the effective channel even when the scalar signal fading is kept unchanged \cite{Qi2015,Soh2015,Marie2017,Laudenbach2019}.

\begin{figure}[htbp]
    \centering
    \includegraphics[width=.4\textwidth]{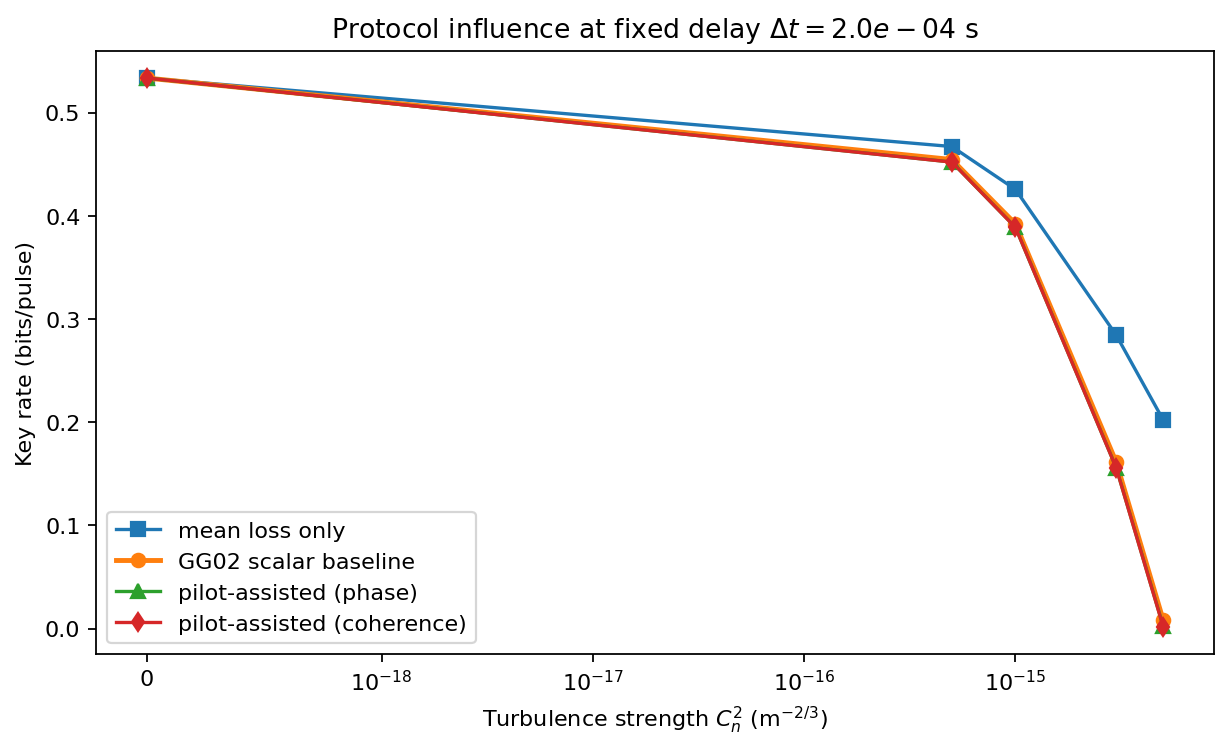}
    \caption{Comparison between the scalar GG02 baseline and pilot-assisted effective models at fixed signal-pilot delay. The pilot-assisted curves lie below the scalar baseline, with a degradation that becomes increasingly significant in the stronger turbulence regime.}
    \label{fig:protocol-keyrates}
\end{figure}

An important point is that the pilot-assisted degradation is already visible at a delay for which the signal and pilot remain strongly correlated. Complete decorrelation is therefore not required for protocol dependence to appear. A modest but finite mismatch between the signal and the transmitted reference can already produce a measurable penalty in the effective key rate, especially near threshold. This behavior is summarized in Fig.~\ref{fig:protocol-keyrates}.

\subsection{Differential decoherence observables}
\label{sec:differential-decoherence}

To understand the optical origin of the additional pilot-assisted noise, it is useful to examine the differential observables introduced in Section~\ref{sec:signal-pilot-model}. The two most informative quantities are the relative phase
\[
\Delta\phi=\arg(a_{\mathrm{s}}a_{\mathrm{p}}^*)
\]
and the complex coherence indicator
\[
\nu=
\frac{\left|\mathbb{E}[a_{\mathrm{p}}^*a_{\mathrm{s}}]\right|}
{\sqrt{\mathbb{E}[|a_{\mathrm{s}}|^2]\mathbb{E}[|a_{\mathrm{p}}|^2]}}.
\]

In the weak-turbulence regime, the relative phase remains narrowly concentrated around zero and the coherence indicator stays close to unity. In this regime, the pilot-assisted excess-noise term is only a small correction to the scalar fading contribution. The transmitted reference therefore remains a good proxy for the signal phase and post-coupling optical mode \cite{Qi2015,Soh2015,Marie2017,Laudenbach2019}.

As turbulence increases, however, the relative phase becomes more dispersed and the signal-pilot coherence decreases. Although these changes may remain modest in absolute terms, they enter the effective reference-noise closures in a way that can substantially reduce the secret key rate, especially when the scalar baseline is already close to threshold. This explains why the pilot-assisted degradation is most visible in the stronger turbulence cases: once the scalar channel is marginal, even a moderate additional reference-noise term can become decisive.

The same conclusion is supported by the signal-pilot coupling correlations. At fixed delay, the signal and pilot couplings remain strongly correlated, which confirms that they still probe closely related optical paths. Nevertheless, the correlation is not perfect, and the residual mismatch is precisely what the pilot-assisted effective model converts into excess noise. The relevant phenomenon is therefore not complete decorrelation, but a progressive loss of differential coherence between the signal and the transmitted reference. A supplementary signal-pilot diagnostic is reported in Appendix~\ref{app:pilot-diagnostics}.

The main text focuses on the phase-reference closure, because the resulting observable \(\sigma_{\phi,\mathrm{eff}}^2\) gives the most stable delay-dependent trend. The coherence-based closure based on \(\nu\) is retained as a useful diagnostic of differential amplitude--phase mismatch, but it is more sensitive to small variations near \(\nu=1\) and to the nonlinear key-rate threshold. Its delay-dependent behavior is therefore reported in Appendix~\ref{app:coherence-delay-diagnostic} rather than used as the primary evidence in the main text.

\subsection{Delay scan and protocol fragility}
\label{sec:delay-scan}

The clearest evidence for protocol dependence is obtained by scanning the signal-pilot delay \(\Delta t\) at fixed turbulence strength. This procedure isolates the effect of frozen-flow decorrelation and shows explicitly how a transmitted reference can cease to track the signal as the delay increases. It is also the most direct way to reveal the existence of a protocol-specific fragility scale that is absent from scalar GG02 descriptions \cite{Taylor1938,Andrews2005,Schmidt2010,Qi2015,Soh2015,Marie2017}.

The delay scan is performed at \(C_n^2=3.0\times10^{-15}\,\mathrm{m}^{-2/3}\), using 300 turbulent realizations per delay and per seed, five independent seeds, and a Fourier sub-pixel implementation of the frozen-flow shift. For each seed, the same signal realizations are reused for all delays, so that the scalar GG02 baseline remains independent of \(\Delta t\). The plotted uncertainty bars correspond to the standard error across independent seeds.

\begin{figure}[htbp]
    \centering
    \includegraphics[width=.42\textwidth]{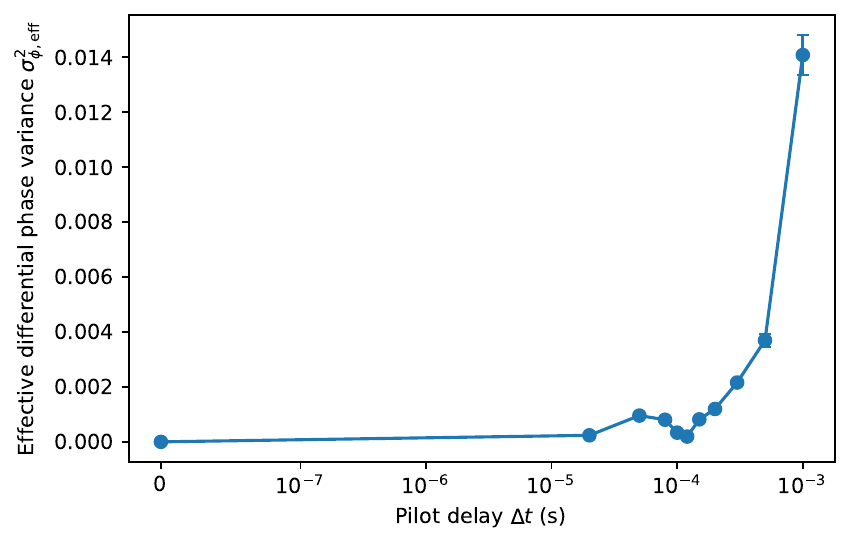}
    \caption{Delay scan at \(C_n^2=3.0\times10^{-15}\,\mathrm{m}^{-2/3}\). The effective differential phase variance \(\sigma_{\phi,\mathrm{eff}}^2\) is estimated from five independent seeds, with 300 turbulent realizations per delay and per seed. The frozen-flow shift is implemented by a Fourier sub-pixel translation of the phase screens.}
    \label{fig:delay-phase-variance}
\end{figure}

The results show that the scalar GG02 baseline is essentially flat as a function of delay, as expected: changing the pilot delay does not change the statistics of the signal coupling itself. By contrast, the phase-reference observable is delay dependent. The effective differential phase variance remains very small at short delays, then grows as the frozen-flow shift becomes large enough to create a measurable signal--pilot mismatch. This behavior is shown in Fig.~\ref{fig:delay-phase-variance}.

\begin{figure}[htbp]
    \centering
    \includegraphics[width=.42\textwidth]{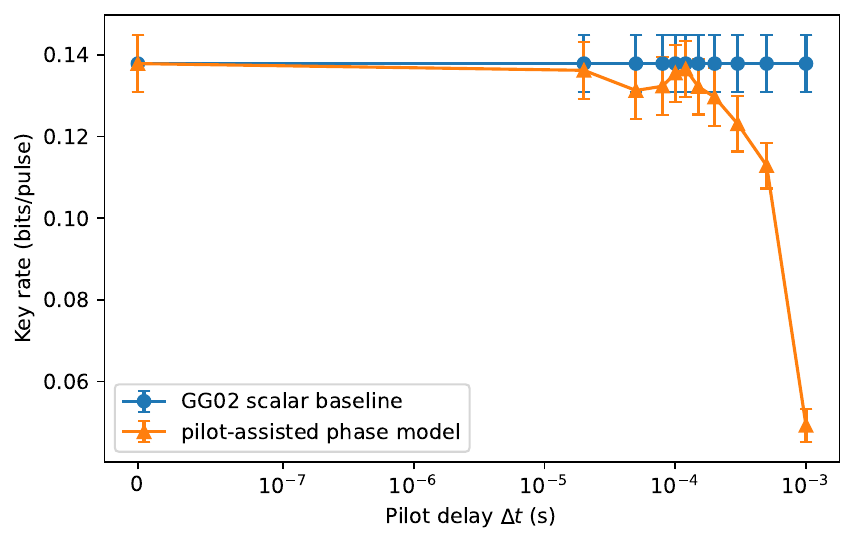}
    \caption{Delay dependence of the effective key rate. The scalar GG02 baseline remains essentially independent of the signal-pilot delay, whereas the pilot-assisted phase model degrades once the differential phase variance becomes non-negligible. Error bars show the standard error over five independent seeds.}
    \label{fig:delay-scan}
\end{figure}

The corresponding key-rate comparison is shown in Fig.~\ref{fig:delay-scan}. The scalar baseline remains stable because it depends only on the signal coupling statistics. The pilot-assisted phase model, however, receives the additional contribution \(\xi_{\mathrm{ref}}^{(\phi)}=V_A\sigma_{\phi,\mathrm{eff}}^2\). It therefore decreases when the differential phase variance becomes appreciable. A local drop-and-recovery feature observed in less averaged diagnostics is not used as a physical conclusion. It is replaced by the more conservative statement supported by the stabilized scan: the robust effect is a delay-dependent phase-reference penalty, while the detailed behavior close to threshold is sensitive to the chosen effective reference-noise closure.

This is the main physical result of the delay scan. The pilot-assisted degradation is not a fixed offset, but a controlled effect driven by differential signal-pilot mismatch. The transmitted reference therefore introduces a new fragility scale that is absent from the scalar GG02 description. A model based solely on the signal coupling law is blind to this scale, whereas a pilot-assisted architecture becomes explicitly sensitive to it. This behavior is summarized in Figs.~\ref{fig:delay-phase-variance} and \ref{fig:delay-scan}.

\subsection{When the protocol becomes physically relevant}
\label{sec:protocol-interpretation}

The results of this section allow a simple interpretation of protocol dependence in the present setting. For GG02, once the full optical propagation has been reduced to a realistic scalar law for the effective signal coupling, the downstream effective channel can be described to good accuracy by that scalar law alone. In that case, no additional optical observable is required beyond the signal-coupling statistics.

By contrast, in the pilot-assisted architecture, the protocol itself creates a second optical degree of freedom, namely the transmitted reference. As soon as this reference is no longer perfectly locked to the signal, an additional observable becomes necessary to describe the effective channel. In the main analysis, this role is played by the relative-phase variance, while the complex coherence indicator is treated as a supplementary diagnostic. The effective channel is therefore no longer specified by the scalar signal-coupling law alone \cite{Qi2015,Soh2015,Marie2017,Laudenbach2019}.

This statement should be interpreted at the level of the effective physical model used here. The present results do not imply that every transmitted-reference protocol requires a fully non-scalar microscopic channel model at the level of a complete security proof. They do show, however, that within the effective framework adopted here, the scalar signal-coupling law is no longer sufficient to account for the observed key-rate degradation. A second observable, measuring signal-pilot differential mismatch, becomes necessary. This is the main physical distinction between the scalar GG02 baseline and the transmitted-reference setting considered in this work.

\section{Conclusion and outlook}
\label{sec:conclusion}

We have studied free-space continuous-variable quantum key distribution with single-mode fiber reception under atmospheric turbulence, with the aim of identifying which receiver-side optical observables are actually needed for effective key-rate modeling.

For the GG02 baseline, the main result is that a description based only on mean loss is generally insufficient and can substantially overestimate the secret key rate. By contrast, once turbulent propagation and SMF reception are modeled explicitly, the downstream effective channel is accurately captured, within the Gaussian closure considered here, by a realistic scalar law for the effective signal coupling. In this sense, the full optical model remains essential upstream because it generates the correct coupling statistics, but the downstream GG02 description is largely governed by a scalar receiver-side observable.

For the pilot-assisted architecture, the situation is different. In that case, the scalar signal-coupling law alone is no longer sufficient to explain the effective key-rate degradation. In addition to scalar fading, the atmosphere induces a differential signal-pilot mismatch that appears after propagation and single-mode projection. In the robust delay scan, this effect is most cleanly captured by the growth of the relative-phase variance, while the coherence-based indicator provides a supplementary diagnostic of amplitude--phase mismatch. The additional contribution behaves as a protocol-dependent reference-noise penalty and becomes increasingly important when the transmitted reference no longer tracks the signal accurately.

The overall physical picture is therefore simple. For GG02 with SMF reception, the key requirement is an accurate effective law for the signal coupling. For transmitted-reference protocols, one must additionally account for signal-pilot differential reference mismatch. The present work is intended as an effective physical modeling framework rather than as a complete security treatment of realistic pilot-assisted CV-QKD. A natural continuation would be to combine the optical observables identified here with more refined security analyses and with more detailed receiver-level models for transmitted-reference architectures.

\appendix

\section{Additional GG02 control results}
\label{app:additional-controls}

For completeness, we report here an additional control comparison supporting the interpretation of the GG02 results discussed in the main text. The figure below compares the full optical model with factorized control models in which the modal-purity contribution is either frozen at its mean value or decorrelated from the aperture-transmission factor. These controls confirm that the full optical propagation is important not only because it generates fluctuations of the effective coupling, but also because it builds the physically relevant joint structure between aperture transmission and modal purity.

\begin{figure}[htbp]
    \centering
    \includegraphics[width=.4\textwidth]{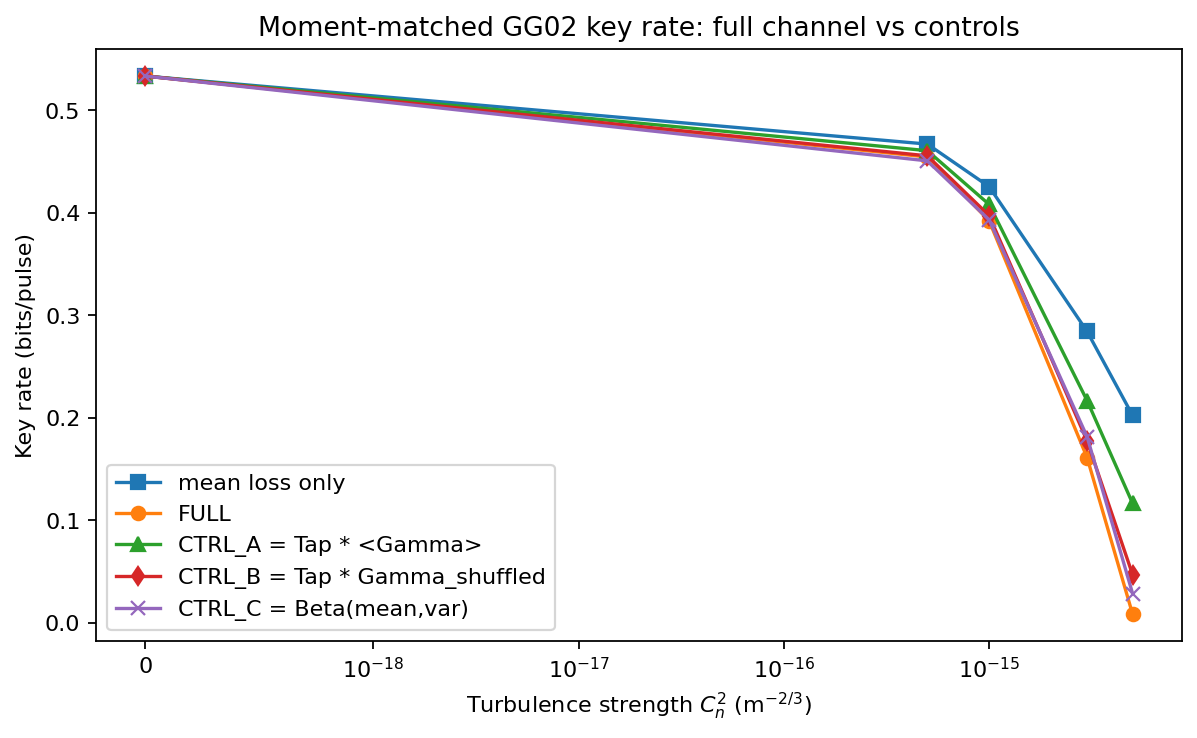}
    \caption{Comparison between the full optical model and factorized control models in the GG02 setting. Freezing the modal-purity factor or destroying its empirical correlation with the aperture-transmission factor leads to systematically more optimistic key-rate estimates, especially in the moderate and strong turbulence regimes.}
    \label{fig:control-keyrates-appendix}
\end{figure}

\section{Supplementary pilot-assisted diagnostics}
\label{app:pilot-diagnostics}

We also include an additional diagnostic for the pilot-assisted setting. While the main text focuses on the resulting key-rate degradation and on the delay dependence of the effective models, the figure below provides a more direct view of the underlying signal--pilot mismatch. It illustrates how the transmitted reference departs from the signal at the level of coupled amplitudes and relative phase, thereby supporting the interpretation of the pilot-assisted excess-noise term as a physically differential decoherence effect.

\begin{figure}[htbp]
    \centering
    \includegraphics[width=.4\textwidth]{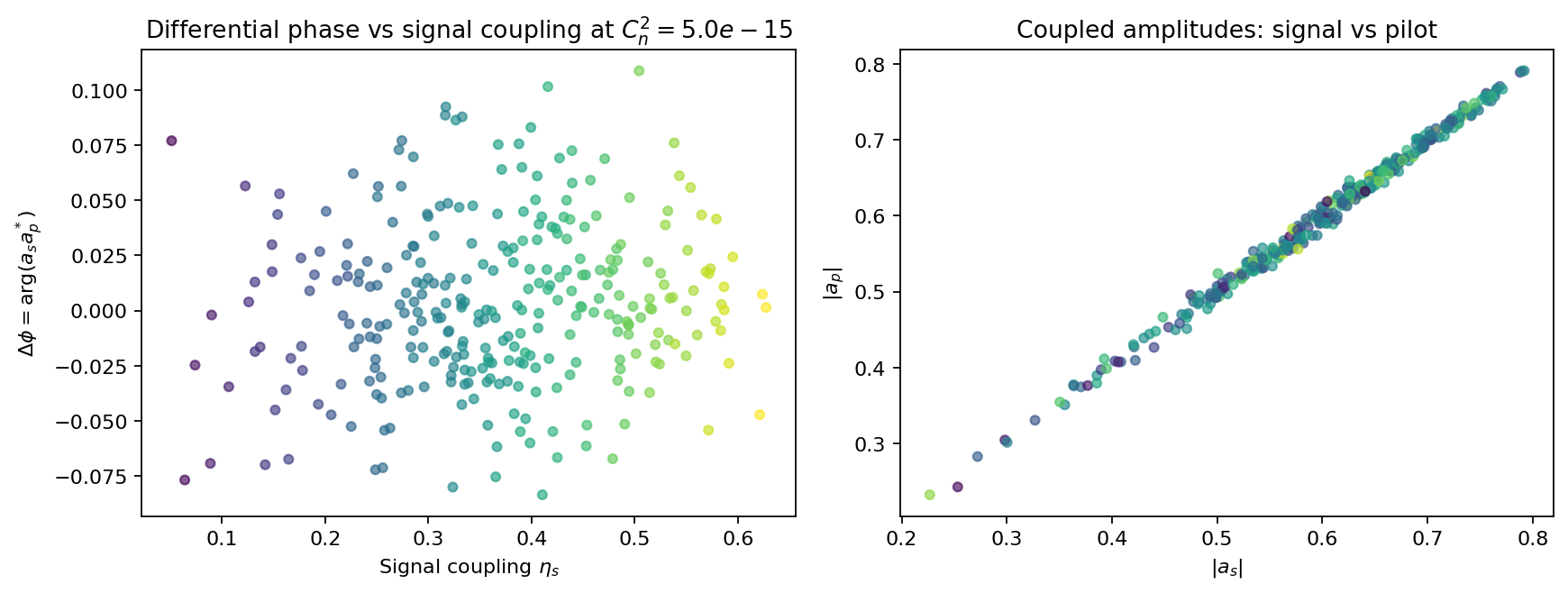}
    \caption{Supplementary pilot-assisted diagnostic showing the signal--pilot mismatch at the level of coupled amplitudes and relative phase. This figure supports the interpretation of the pilot-assisted degradation as a consequence of differential signal--pilot decoherence rather than of scalar signal fading alone.}
    \label{fig:phase-scatter-appendix}
\end{figure}

\section{Coherence-based delay diagnostic}
\label{app:coherence-delay-diagnostic}

The main text uses the phase-reference closure as the primary delay-scan result. For completeness, we also report the coherence-based delay diagnostic obtained from the same multi-seed simulations. This diagnostic is useful because it probes the full complex compatibility of the signal and pilot coupled amplitudes. However, the corresponding excess-noise model \(\xi_{\mathrm{ref}}^{(\nu)}=2V_A(1-\nu)\) can strongly amplify small fluctuations when \(\nu\) is close to unity and the key rate is near threshold. We therefore interpret this curve as a sensitivity diagnostic rather than as the main quantitative delay law.

\begin{figure}[htbp]
    \centering
    \includegraphics[width=.42\textwidth]{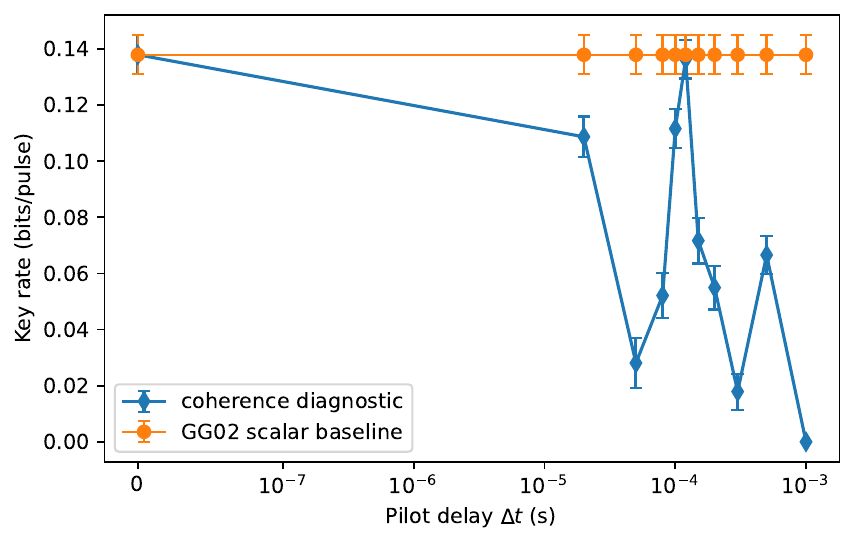}
    \caption{Supplementary coherence-based delay diagnostic. The curve illustrates the sensitivity of the coherence closure to small variations of the complex signal-pilot compatibility and to key-rate threshold effects. For this reason, the main text focuses on the phase-reference delay scan.}
    \label{fig:coherence-delay-diagnostic}
\end{figure}

\section{Convergence diagnostics}
\label{app:convergence}

For completeness, we report here a representative convergence diagnostic for the Monte Carlo estimation of the effective channel parameters and key-rate quantities. The figure below shows the stabilization of representative observables as the number of turbulent realizations increases. It confirms that the numerical comparisons reported in the main text are not artifacts of insufficient sampling.

\begin{figure}[htbp]
    \centering
    \includegraphics[width=.4\textwidth]{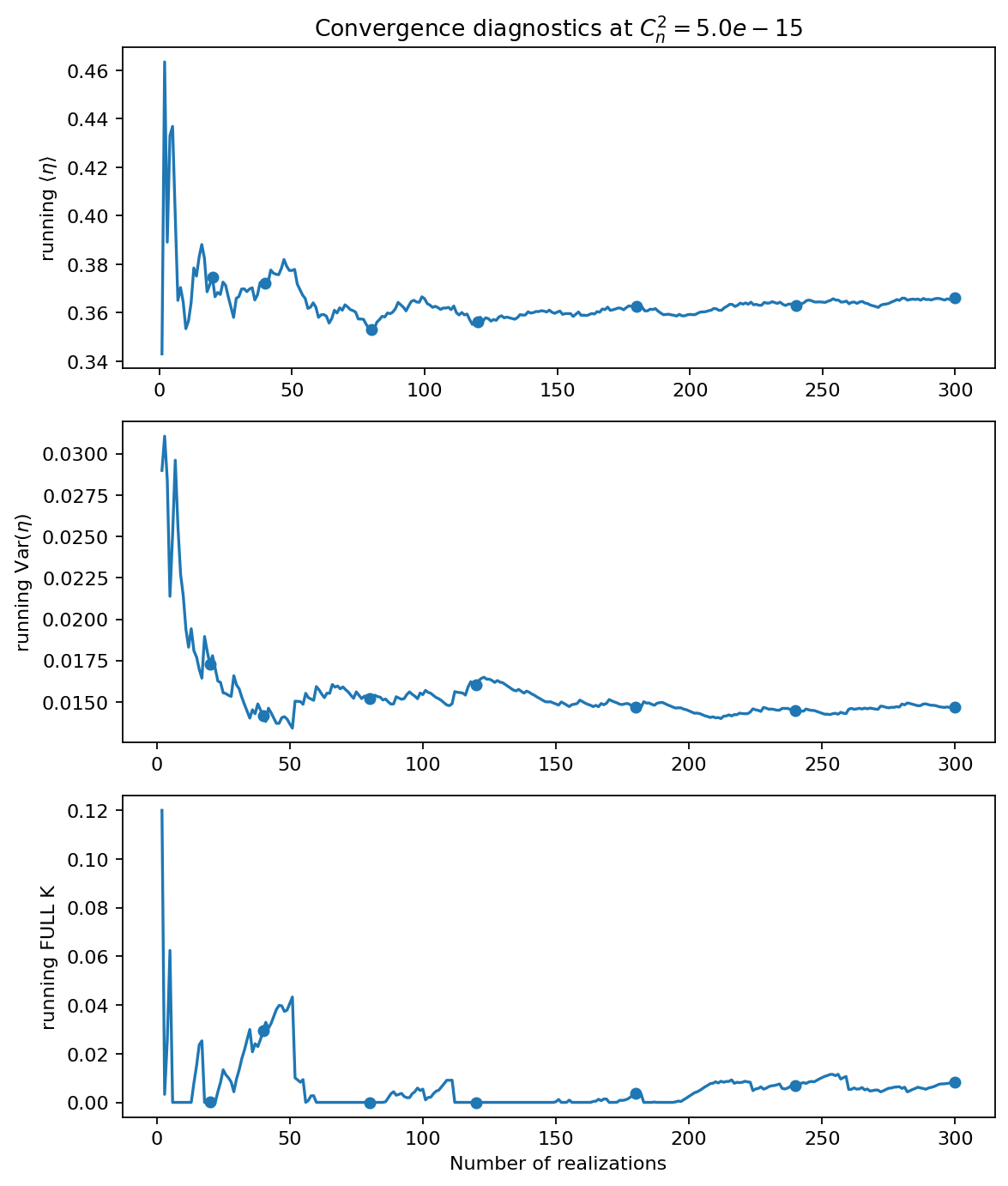}
    \caption{Representative convergence diagnostics for the Monte Carlo estimation of effective optical and key-rate observables as a function of the number of turbulent realizations.}
    \label{fig:convergence-appendix}
\end{figure}

\end{document}